\documentclass{article}
\usepackage{arxiv}
\usepackage[utf8]{inputenc}
\usepackage[T1]{fontenc}
\usepackage{lmodern}
\usepackage{hyperref}
\usepackage{url}
\usepackage{booktabs}
\usepackage{amsmath,amssymb,amsfonts,amsthm}
\usepackage{microtype}
\usepackage{graphicx}
\usepackage{natbib}
\usepackage{doi}
\usepackage{enumitem}
\usepackage{array}
\usepackage{float}
\usepackage{longtable}

\newcommand{\ioxtversion}{0.1.0}
\newtheorem{definition}{Definition}

\setlist{nosep,leftmargin=*}
\title{IoXT: The Internet of Explainable Things\\
\large Why Explainability in IoT Requires a New System-Level Paradigm and Protocol Design}

\author{
\href{https://orcid.org/0000-0002-5930-8814}{\includegraphics[scale=0.055]{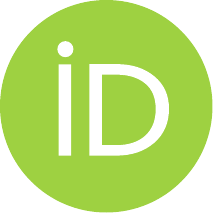}\hspace{1mm}Michael Georgiades}\\
Department of Computer Science\\
Neapolis University Pafos, Paphos, Cyprus\\
\texttt{m.georgiades@nup.ac.cy}
}

\date{Version \ioxtversion\ -- August 2026}
\renewcommand{\shorttitle}{IoXT: The Internet of Explainable Things}

\hypersetup{
  hidelinks,
  pdftitle={IoXT: The Internet of Explainable Things - Why Explainability in IoT Requires a New System-Level Paradigm and Protocol Design},
  pdfauthor={Michael Georgiades},
  pdfsubject={Internet of Explainable Things; explainability-by-design; cross-layer explainability; end-to-end traceability; IoT architecture; protocol design},
  pdfkeywords={IoXT, Internet of Explainable Things, IoT, XAI, explainability-by-design, cross-layer explainability, traceability, protocol design, XTP}
}

\begin{document}
\maketitle

\begin{abstract}
The Internet of Things (IoT) increasingly combines sensing, communication, artificial intelligence (AI), decision-making, and actuation. In many domains, sensor observations are processed by edge or cloud intelligence to select actions that configure or control actuators; where actuation changes the environment, later observations may also be affected. Existing explainable AI (XAI) methods can explain model predictions, but they do not by themselves explain the end-to-end path from sensed evidence to physical action.

This paper introduces the \emph{Internet of Explainable Things (IoXT)}, a system-level paradigm that makes explainability an architectural property of intelligent IoT. Its novelty is \emph{explainability-by-design} across the sensing--communication--intelligence--decision--actuation path. IoXT derives requirements and design principles for identity and addressability, temporal fidelity, cross-layer provenance, bidirectional traceability, streaming and incident-time evidence, protocol-semantic continuity, security and privacy, lifecycle continuity, and conformance. We formalize timestamped provenance graphs, sensor participation, trace completeness, cross-layer coverage, reconstruction latency, and \emph{No Orphan Actuation}: a consequential action must remain traceable to the authorizing decision and source evidence, or be explicitly marked degraded or non-conformant.

IoXT prescribes neither a particular XAI method nor a wire protocol. Instead, it defines the conditions under which model explanations remain connected to real sensor inputs, communication history, decisions, actuator execution, and outcomes. \emph{Explainability Telemetry Protocols (XTPs)} are introduced as a protocol category for preserving these semantics across heterogeneous IoT systems, together with \emph{IoXT-ready} and \emph{IoXT-conformant} assurance concepts.
\end{abstract}


\clearpage
\section{Introduction}
IoT has moved beyond isolated sensing toward distributed systems that sense, communicate, infer, decide, and act. A useful progression is
\begin{equation}
\mathrm{IoT}\;\longrightarrow\;\mathrm{AIoT}\;\longrightarrow\;\mathrm{IoXT},
\label{eq:progression}
\end{equation}
where the proposed \emph{Internet of Explainable Things (IoXT)} denotes an architectural stage in which intelligent Things must also preserve the evidence required to explain their behaviour end to end.

There is already a substantial XAI-for-IoT literature \citep{kok2023explainable,jagatheesaperumal2022explainable,moss2025explainable,mohammad2025strategies,ambritta2024explainable}. IoXT does not replace model-level XAI. Methods such as LIME and SHAP can be used as examples of local explanation and feature attribution \citep{ribeiro2016lime,lundberg2017unified}; feature-importance methods can rank influential inputs \citep{altmann2010permutation}; and intrinsic models such as Self-Explaining Neural Networks (SENN) can expose concepts and relevance scores during prediction \citep{alvarezmelis2018senn}. C-SENN illustrates how concept learning can be further differentiated in more complex environments \citep{sawada2022csenn}. These techniques can provide the \emph{model} part of an IoXT explanation, but they do not by themselves establish which physical sensor produced the evidence, how that evidence travelled and changed, which contemporaneous model/policy state consumed it, or which actuator executed the resulting command.

That distinction is especially important in IoT because both the \emph{data} and the \emph{communication of the data} may be explanation-relevant. Cross-layer IoMT work shows that preserving packet-, application- and sensing-related semantics can reveal explanatory evidence lost by single-layer abstractions \citep{georgiades2025explainable}. More broadly, provenance and distributed tracing research shows the value of retaining data origin, processing history and cross-component relationships \citep{hu2020survey,lomotey2018traceability,fonseca2007x}; XAI-for-cyber-physical-systems research similarly recognizes that physical, sensing, control, safety and security context complicate explanation \citep{hoenig2024explainable}.

The systems challenge is reinforced by the broader IoT literature. IoT commonly joins sensors, informational processing and actuators so that informational failures can have physical consequences \citep{allhoff2018internet}. Ethical and legal work highlights pervasive sensing, cloud processing, actuation, transparency, privacy, safety, trust and accountability \citep{baldini2018ethical,tzafestas2018ethics,popescul2013internet,abobakr2017iot,berman2017social,karale2021challenges,Atlam2020,wachter2018gdpr}. Management and orchestration research adds identity, monitoring, configuration, fault handling, heterogeneous lifecycle state, service migration and edge--cloud coordination \citep{aboubakar2022review,sinche2019survey,wen2017fog,shahid2026iot}. Governance research further emphasizes transparency, accountability, interoperability and oversight \citep{weber2013internet}. Together, these strands motivate the core IoXT claim: \textbf{if explainability is required across a distributed, time-varying sensing--intelligence--actuation path, it must be engineered into the architecture from the beginning.}

Explainability-by-design research provides an important methodological precedent for building explanatory capability into systems rather than treating it only as a post-hoc activity \citep{huynh2022methodology}. Likewise, streaming-data research shows that model and data distributions can change over time, so explanations detached from the event-time system state can be misleading \citep{lu2018learning,hu2024explainable}. IoXT therefore focuses on what must be captured while the system is operating.

The paper makes five contributions. First, it explains why IoT explainability is a system problem rather than only a model problem. Second, it defines a requirements- and design-principles-first IoXT architecture. Third, it formalizes end-to-end traceability, incident reconstruction, no-orphan actuation and measurable system properties. Fourth, it motivates XTP as a category of explanation-telemetry mechanisms without prescribing a single protocol. Fifth, it outlines assurance concepts and research questions for future standardization and domain-specific certification.

\clearpage
\section{Why Explainability in IoT Is Different}
\subsection{The Common Sensor--Intelligence--Actuator Pattern}
A conventional XAI question often asks why model $M$ produced output $\hat{y}$ for input $x$. Across domains such as smart buildings, industrial automation, energy, mobility, agriculture and healthcare, a common operational pattern is that sensed state informs intelligence, and intelligence selects a configuration or action for one or more actuators. A more complete operational path is
\begin{equation}
 e_t \xrightarrow{S} x_t \xrightarrow{N_t} z_t \xrightarrow{M_t} \hat{y}_t
 \xrightarrow{\Pi_t} a_t \xrightarrow{A_t} e_{t+\Delta},
\label{eq:loop}
\end{equation}
where $e_t$ is the relevant physical or operational environment at time $t$, $S$ denotes sensing, $N_t$ communication and preprocessing, $M_t$ the model active at that time, $\Pi_t$ the decision/control policy, $a_t$ the selected action or command, and $A_t$ the actuator state. Equation~\eqref{eq:loop} represents a common sensor--intelligence--actuator scenario rather than a requirement that every IoT system be closed-loop. Some IoT systems stop at sensing, analytics or human decision support. Where actuation is present, however, the action may alter the environment and therefore influence later observations.

For example, a temperature sensor may produce reading $T_t$, intelligence may select a fan-speed command $v_t$, and the fan may alter the temperature subsequently observed:
\begin{equation}
 v_t = \Pi_t\!\left(M_t\!\left(N_t(T_t)\right),C_t\right),
 \qquad
 T_{t+\Delta}=F(T_t,v_t,d_t),
\label{eq:fan}
\end{equation}
where $C_t$ captures relevant decision context and $d_t$ represents external disturbances. If the selected action is inappropriate, the command is not executed, or the resulting physical response is abnormal, an investigator should be able to move backwards from the action and outcome to the decision, model state, communication/transformation history, and the specific timestamped sensor observations that served as inputs.

IoXT must therefore answer more than ``Why did the model predict $\hat{y}_t$?'' It must support questions such as: Which sensors and readings informed the selected action? At what times were they observed? Were the observations delayed, aggregated or transformed? Which model and policy versions were active? Which actuator received and executed the command? What happened afterwards? These are system explanations. The same traceability requirement is useful even when there is no physical feedback loop: the action or recommendation should still remain connected to the evidence that produced it.

\subsection{Data, Communication and Cross-Layer Context Are All Evidence}
IoXT treats the explanation scope conceptually as
\begin{equation}
\mathcal{E}_{\mathrm{IoXT}}
=\mathcal{E}_{\mathrm{data}}
\cup\mathcal{E}_{\mathrm{comm}}
\cup\mathcal{E}_{\mathrm{model}}
\cup\mathcal{E}_{\mathrm{decision}}
\cup\mathcal{E}_{\mathrm{action}}.
\label{eq:scope}
\end{equation}
The communication component matters because packet loss, buffering, aggregation, protocol translation, gateway filtering, unit conversion, duplication, reordering and edge fusion can change what the model effectively observes. Two numerically similar model inputs can therefore have different provenance and reliability.

The cross-layer IoMT framework of \citet{georgiades2025explainable} is an instructive example: it preserves packet-, application- and sensing-related semantics rather than explaining an intrusion decision only through aggregated flow features. IoXT generalizes this idea. When evidence is distributed across layers, explanation must preserve the relationships between those layers, even if constrained components expose only compact references rather than full internal state.

\subsection{Time Makes Explanation Ephemeral}
The explanation of an IoT action is time-indexed because sensing, communication, model, policy, orchestration and actuator states evolve. We represent the contemporaneous explanation state as
\begin{equation}
\mathcal{X}_t = \{S_t,N_t,M_t,\Pi_t,A_t,C_t\},
\label{eq:state}
\end{equation}
where $C_t$ denotes relevant context such as configuration, deployment, environmental and governance state. If a model is retrained after an incident, a gateway is reconfigured or a sensor is recalibrated, the current state cannot be assumed to explain the historical action. A timestamp is therefore not auxiliary metadata; it is part of the explanation.

IoXT distinguishes three modes: \emph{streaming explainability}, in which compact explanation/provenance state is generated or referenced while data are flowing; \emph{incident-time explainability}, in which an event triggers immediate preservation and traversal; and \emph{retrospective audit}, in which richer offline analysis uses contemporaneous evidence captured during operation. This is particularly important in streaming environments subject to drift and model evolution \citep{lu2018learning,hu2024explainable}.

\subsection{Distribution Makes Explainability an Architecture and Networking Problem}
Explanation state may be distributed across sensors, gateways, brokers, edge nodes, cloud models, policy engines and actuators owned by different vendors. A later XAI algorithm cannot reconstruct a relationship that was never represented in device state, message metadata or a provenance record. This leads to a simple design rule:
\begin{quote}
\textbf{What must be explainable must also be identifiable, timestamped, linkable and retrievable.}
\end{quote}
The network does not need to carry a full explanation in every packet. It must, however, preserve enough semantics to join an actuator action to the decision, model explanation, communication/transformation history and sensor evidence that influenced it.

\subsection{Retrofitting After the Incident Is Too Late}
Retrofitting remains possible for legacy systems, but only for evidence that the system is capable of capturing. If a deployment never recorded which sensor produced a value, which preprocessing step modified it, which model version consumed it or which decision authorized an actuator command, those links cannot be recreated with certainty after an accident. IoXT therefore extends by-design thinking to explainability itself: the required hooks must be introduced during architecture, enrolment and commissioning, before the evidence is needed \citep{huynh2022methodology,baldini2018ethical}.

\clearpage
\section{IoXT Requirements and Design Principles}
IoXT is deliberately requirements-first. The requirements state \emph{what} a system must preserve; design principles guide \emph{how} future architectures and standards should satisfy them.

\subsection{Core System Requirements}
\begin{table}[H]
\centering
\small
\caption{Core IoXT system requirements.}
\label{tab:req}
\begin{tabular}{p{0.09\linewidth} p{0.83\linewidth}}
\toprule
\textbf{ID} & \textbf{Requirement} \\
\midrule
R1 & Explanation-relevant Things, logical components and events shall be uniquely identifiable and queryable. \\
R2 & Observation, processing, decision, command and execution events shall carry timestamps and, where necessary, clock-quality or uncertainty metadata. \\
R3 & The system shall preserve end-to-end lineage from sensed evidence through communication, transformation, intelligence and decision to actuation. \\
R4 & Trace-back from action to source evidence and trace-forward from suspect evidence to downstream actions shall be supported. \\
R5 & Explanation state shall be available during streaming operation and preservable at incident time, not created only from an offline dataset afterwards. \\
R6 & Model-level explanations, when available, shall remain linked to the sensor-derived inputs and contemporaneous model/policy state that produced the decision. \\
R7 & Protocol translation, gateway processing, service migration and orchestration shall preserve required trace semantics or explicitly report degradation. \\
R8 & Explanation evidence shall be protected for integrity, authenticity, confidentiality/privacy, authorization and retention. \\
R9 & Constrained Things may use references, hashes, summaries or delegated storage, provided that the end-to-end trace remains verifiable. \\
R10 & A critical actuation shall either have a complete required trace or be explicitly marked degraded/non-conformant. \\
\bottomrule
\end{tabular}
\end{table}

\subsection{Design Principles}
\paragraph{P1: Explainability-by-design.} Explanation hooks are introduced during architecture, enrolment and commissioning. Retrofitting is a migration strategy, not a substitute for contemporaneous evidence capture.

\paragraph{P2: Identity and addressability.} Every Thing, event or logical component that can influence a consequential decision needs a stable, queryable identity in the explanation path. This identity need not be an IP address; it may represent a sensor channel, transformation, model instance, policy, action or outcome. Machine-readable Thing descriptions provide useful precedent for structured identity and metadata \citep{kaebisch2023web}.

\paragraph{P3: Temporal fidelity and tracking.} Tracking maintains evolving identity, placement, version and state during operation; traceability reconstructs the relationships that produced a particular action or were influenced by a particular source. The two are complementary.

\paragraph{P4: End-to-end, cross-layer provenance.} The trace must survive sensing, network/protocol processing, gateways, edge/cloud transformations, model inference, policy logic and actuation. General provenance models such as W3C PROV provide useful vocabulary for entities, activities and derivations \citep{w3c2013provdm}, but IoXT additionally requires communication and physical-actuation semantics.

\paragraph{P5: Bidirectional traceability.} Trace-back starts from an action or outcome and moves toward the source evidence. Trace-forward starts from a suspect sensor, message or transformation and identifies downstream decisions and actions.

\paragraph{P6: Streaming and incident availability.} The system continuously retains or references the minimum evidence required to reconstruct a live trace. A serious event can freeze or attest the relevant window before transient state is lost.

\paragraph{P7: Technology-neutral protocol semantics.} IoXT does not require one transport. It requires that identity, time, correlation, parentage, version/context and integrity semantics survive whatever protocols and gateways carry them.

\paragraph{P8: Protected and governable evidence.} Explanation state can itself be sensitive and attackable. Security, privacy, management, orchestration and governance are orthogonal requirements, not post-processing concerns. The EU AI Act and NIST AI RMF provide relevant external motivation for logging, traceability, transparency and lifecycle risk management without defining IoXT itself \citep{eu2024aiact,nist2023airmf}.

\clearpage
\section{A Multi-Plane IoXT Architecture}
\begin{figure}[H]
\centering
\includegraphics[width=0.98\textwidth]{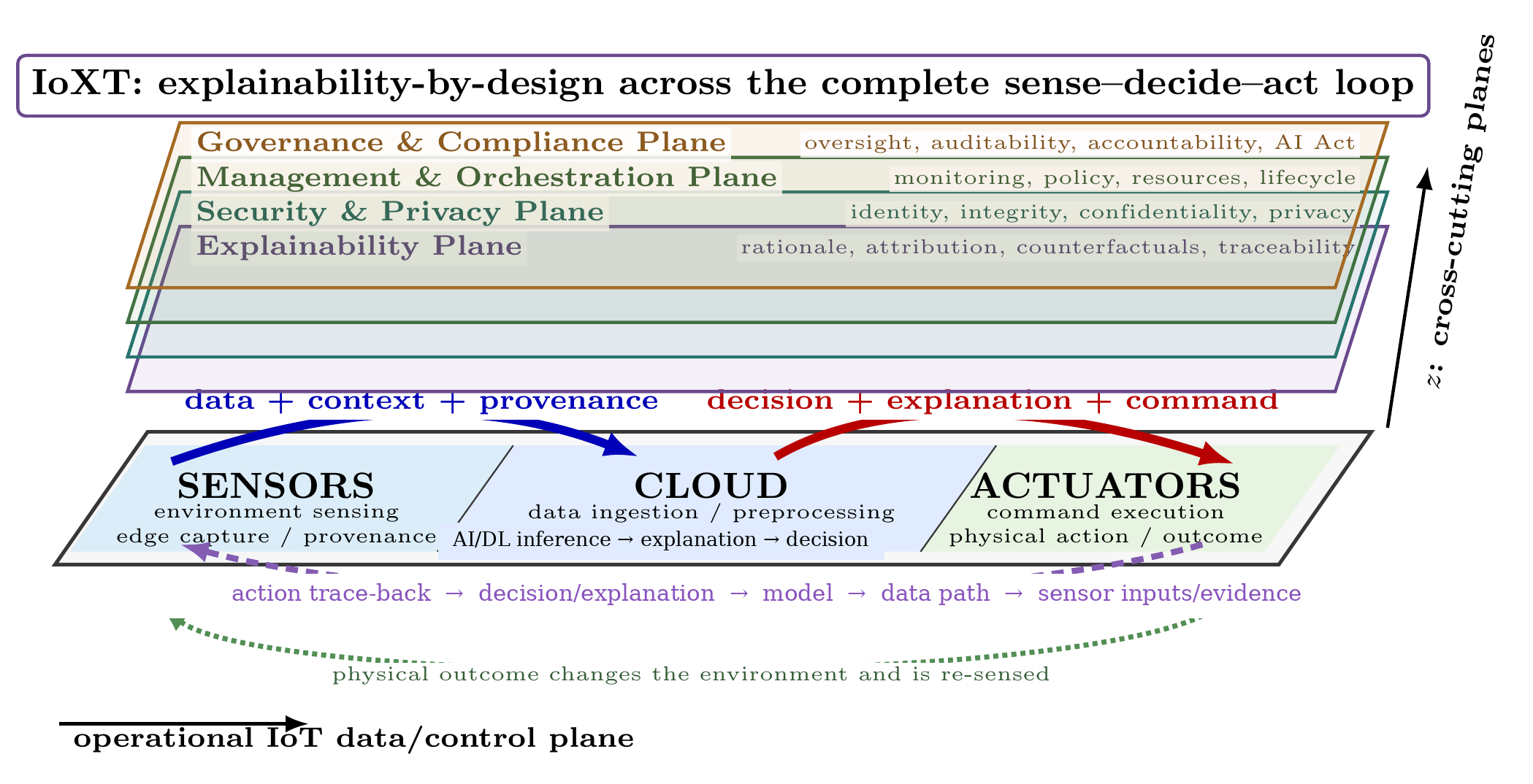}
\caption{IoXT multi-plane architecture. The operational plane follows the common Sensors--Cloud/Intelligence--Actuators path; when actuation changes the environment, feedback creates a subsequent sensing cycle. The reverse trace links an actuator action to the decision, model explanation, communication/data path and timestamped sensor evidence that produced it.}
\label{fig:ioxt}
\end{figure}

Figure~\ref{fig:ioxt} places the common Sensors--Cloud/Intelligence--Actuators path in the horizontal plane. Four cross-cutting planes intersect the whole path: explainability; security and privacy; management and orchestration; and governance and compliance. Where actuator actions influence the environment, the feedback shown in the figure represents the next sensing cycle; it is a common cyber-physical case, not a requirement for every IoT deployment. The main architectural claim is not that these concerns are new individually. It is that explainability must be elevated to the same system-level status.

Within the intelligence stage, the model may use any suitable explanation mechanism. For example, LIME and SHAP can provide local explanations or feature attributions \citep{ribeiro2016lime,lundberg2017unified}; permutation feature importance can rank influential variables \citep{altmann2010permutation}; SENN can expose concepts and relevance scores intrinsically \citep{alvarezmelis2018senn}; and C-SENN can encourage more differentiated concepts in complex environments \citep{sawada2022csenn}. IoXT does not prescribe one of these. It requires whichever explanation is produced to remain connected upstream to the sensor-derived evidence and communication path, and downstream to the decision, actuator execution and physical outcome.

\clearpage
\section{Formal Model, Definitions and Metrics}
The following definitions make the IoXT requirements measurable without fixing an implementation.

\begin{definition}[Traceable Thing]
A Thing that can influence a consequential decision must be traceable inside the explanation path for that decision.
\end{definition}
For a sensor, the minimum trace state includes identity, measurement time, calibration/configuration version, and quality/uncertainty where available. For an actuator, it includes identity, command/execution time, decision reference, local execution state and outcome. Gateways and intelligence components preserve the transformations and versioned state connecting the two ends.

\begin{definition}[Timestamped incident provenance graph]
For an incident associated with action $a_t$, let $G_t=(V_t,E_t)$ be the timestamped provenance graph frozen for the relevant interval. A vertex represents an explanation-relevant entity or event; a directed edge represents an explanation-relevant dependency, transformation or authorization relation.
\end{definition}
If $s_i$ denotes a sensor entity or observation, the sensor observations that participated in the action path are
\begin{equation}
\mathcal{S}(a_t)=\{s_i\in V_t\mid s_i\rightsquigarrow a_t\text{ in }G_t\}.
\label{eq:sensorset}
\end{equation}
A reverse query should recover $\mathcal{S}(a_t)$ together with timing, calibration, transformations, communication state, model/policy state, model-level explanation and actuator state.

Conversely, if sensor $s_j$ is later found faulty or compromised over interval $[t_0,t_1]$, the forward impact set is
\begin{equation}
\mathcal{A}(s_j;[t_0,t_1])=
\{a_k\mid s_j\rightsquigarrow a_k\text{ in }G_{[t_0,t_1]}\}.
\label{eq:impactset}
\end{equation}
Equations~\eqref{eq:sensorset} and \eqref{eq:impactset} establish participation and reachability, not automatically legal or counterfactual causation. They provide the time-correct evidence on which causal, engineering or forensic analysis can operate.

\begin{definition}[No Orphan Actuation]
A safety- or policy-relevant action is an \emph{orphan action} when the required trace schema cannot resolve it back to the authorizing decision and source evidence.
\end{definition}
For action $a_t$, let $\Gamma(a_t)$ denote the required provenance links and $\Gamma^{\mathrm{obs}}(a_t)$ the verified links available at audit time. Define trace completeness as
\begin{equation}
C_{\mathrm{trace}}(a_t)=
\frac{|\Gamma^{\mathrm{obs}}(a_t)|}{|\Gamma(a_t)|},
\qquad 0\leq C_{\mathrm{trace}}\leq 1.
\label{eq:ctrace}
\end{equation}
For a declared critical action, an IoXT conformance profile should require $C_{\mathrm{trace}}(a_t)=1$ or an explicit degraded/non-conformant state. The intention is not indefinite retention of all raw sensor traffic; domain profiles may define the minimum evidence, summaries or cryptographic references that must be retained.

For the set $\mathcal{A}_c$ of critical actions, the orphan-actuation rate is
\begin{equation}
R_{\mathrm{orphan}}=
\frac{1}{|\mathcal{A}_c|}
\sum_{a\in\mathcal{A}_c}\mathbf{1}[C_{\mathrm{trace}}(a)<1].
\label{eq:orphan}
\end{equation}
A mature critical deployment should target $R_{\mathrm{orphan}}=0$.

\subsection{Cross-Layer Coverage}
Let $\mathcal{L}^{\mathrm{req}}(a_t)$ be the explanation layers required by a domain profile, for example sensing, communication, preprocessing, model, decision and actuation, and let $\mathcal{L}^{\mathrm{obs}}(a_t)$ be those successfully represented in the reconstructed trace. Define
\begin{equation}
C_{\mathrm{layer}}(a_t)=
\frac{|\mathcal{L}^{\mathrm{obs}}(a_t)|}{|\mathcal{L}^{\mathrm{req}}(a_t)|}.
\label{eq:layer}
\end{equation}
This metric distinguishes a complete system explanation from a faithful explanation of only one layer.

\subsection{Temporal Consistency}
Let $E_t(a_t)\subseteq E_t$ denote the edges in the reconstructed action trace. For an edge $(u,v)$, let $\epsilon_{uv}$ be an allowed temporal tolerance accounting for clock uncertainty and buffering. A temporal-order violation occurs when $t(u)>t(v)+\epsilon_{uv}$. Define
\begin{equation}
V_{\mathrm{time}}(a_t)=
\sum_{(u,v)\in E_t(a_t)}
\mathbf{1}[t(u)>t(v)+\epsilon_{uv}],
\label{eq:timeviol}
\end{equation}
and
\begin{equation}
C_{\mathrm{time}}(a_t)=
1-\frac{V_{\mathrm{time}}(a_t)}{|E_t(a_t)|}.
\label{eq:ctime}
\end{equation}
$C_{\mathrm{time}}=1$ means that all observed causal/processing relationships are temporally consistent within their declared uncertainty. This is preferable to assuming perfect synchronization in heterogeneous IoT deployments.

\subsection{Reconstruction and Explanation Availability Latency}
During an incident, the time required to reconstruct an explanation matters. Define
\begin{equation}
L_{\mathrm{recon}}(a_t)=
 t_{\mathrm{trace\ ready}}-t_{\mathrm{query}}.
\label{eq:latency}
\end{equation}
A separate streaming metric can measure the age of the newest explanation evidence when an action occurs:
\begin{equation}
L_{\mathrm{evidence}}(a_t)=
 t(a_t)-\max_{r_i\in\mathcal{R}(a_t)} t(r_i),
\label{eq:evidenceage}
\end{equation}
where $\mathcal{R}(a_t)$ is the evidence set used to justify the action. Domain profiles can place upper bounds on this quantity when stale evidence is unsafe.

\subsection{Minimum Explanation-Telemetry Envelope}
A compact explanation-telemetry record can be represented generically as
\begin{equation}
r_i=\langle id_i,t_i,\tau_i,P_i,v_i,c_i,q_i,h_i\rangle,
\label{eq:envelope}
\end{equation}
where $id_i$ is the addressable entity/event identity, $t_i$ event time, $\tau_i$ a trace/correlation identifier, $P_i$ parent/input links, $v_i$ version state, $c_i$ operational context, $q_i$ quality/uncertainty state, and $h_i$ integrity or attestation information. A rich model explanation may be referenced rather than carried inline. The purpose of Eq.~\eqref{eq:envelope} is to define minimum semantics, not a mandatory packet format.

\clearpage
\section{Protocol Requirements and the XTP Category}
The need for protocol design follows from distribution: explanation state must cross devices, brokers, gateways, administrative domains and protocol boundaries. IoXT therefore uses \textbf{Explainability Telemetry Protocol (XTP)} as a \emph{category}, not a fixed protocol. An XTP realization may be an extension to an existing IoT protocol, a profile standardizing trace metadata, a gateway-preservation rule set, a companion/sidecar channel, or a new lightweight protocol if existing mechanisms cannot efficiently carry the required semantics.

Let $\mathcal{F}_{\mathrm{req}}$ denote the required explanation semantics on a path, such as identity, event time, correlation, parentage, version/context and integrity. For protocol/gateway stage $k$, let $\mathcal{F}^{(k)}_{\mathrm{obs}}$ be the semantics available after the mapping. A simple semantic-preservation score is
\begin{equation}
C_{\mathrm{sem}}^{(k)}=
\frac{|\mathcal{F}^{(k)}_{\mathrm{obs}}\cap\mathcal{F}_{\mathrm{req}}|}
{|\mathcal{F}_{\mathrm{req}}|}.
\label{eq:csem}
\end{equation}
For a critical trace, an XTP-capable gateway should either maintain $C_{\mathrm{sem}}^{(k)}=1$ for the required profile or explicitly report the missing semantics. Silent trace degradation is incompatible with explainability-by-design.

At minimum, an XTP-capable mechanism should support operations equivalent to: publish a timestamped explanation/provenance record; link an output to parent evidence; query by identity, trace or time; subscribe to live trace events; and preserve/attest an incident interval. The exact wire representation is deliberately left open.

Existing protocols offer design precedents without defining IoXT. SNMP demonstrates remotely addressable management state \citep{rfc3411}; CoAP demonstrates resource-oriented communication and discovery for constrained devices \citep{rfc7252}; and MQTT demonstrates lightweight publish--subscribe messaging for IoT/M2M contexts \citep{mqtt5}. IoXT abstracts the relevant properties -- addressability, constrained operation, telemetry, query, subscription and gateway behaviour -- while adding the explanation semantics required by the end-to-end cyber-physical trace.

\clearpage
\section{Assurance: IoXT-Ready, IoXT-Conformant and Incident Evidence}
\begin{definition}[IoXT-ready Thing]
An \emph{IoXT-ready Thing} exposes sufficient identity, time, version, lineage, quality and integrity state to participate in a verifiable explanation trace for the functions it performs.
\end{definition}

\begin{definition}[IoXT-conformant path]
An \emph{IoXT-conformant path} preserves the required explanation trace across the complete Sensors--Intelligence--Actuators path for declared consequential actions and can demonstrate both trace-back and trace-forward under the applicable domain profile.
\end{definition}

\begin{definition}[IoXT-conformant loop]
An \emph{IoXT-conformant loop} is an IoXT-conformant path in which actuation influences subsequent sensing and the trace remains continuous across successive sensing--decision--action cycles.
\end{definition}

\begin{definition}[IoXT Certified]
\emph{IoXT Certified} is proposed as a future assurance label that could be awarded only after independent testing against an agreed, domain-specific IoXT conformance profile. It is not an existing certification scheme and should not be self-asserted.
\end{definition}

A future conformance test should verify at least: addressable identity; timestamps and clock quality; calibration/configuration/model/policy version state; parent/input lineage; live export or reference of provenance; action-to-decision linkage; incident preservation; reverse and forward traversal; integrity and access control; and explicit reporting of incomplete traces. Equations~\eqref{eq:ctrace}--\eqref{eq:csem} provide an initial measurable core.

\subsection{A Distributed Explainability Flight Recorder}
A particularly useful realization is a distributed cyber-physical \emph{explainability flight recorder}. During normal operation, components emit or reference compact, time-indexed evidence. When an anomaly, unsafe action or accident occurs, the relevant interval is frozen and integrity-protected across participating components. The incident bundle should answer: what did the sensors observe, what was their calibration and quality state, how did the data travel and change, which model and policy were active, what model-level explanation was produced, which actuator acted, and what happened next?

This concept translates long-standing accountability, safety and transparency concerns in IoT \citep{berman2017social,tzafestas2018ethics,karale2021challenges} into a concrete systems requirement. It also supports later causal, engineering, safety or legal analysis without claiming that provenance alone proves causation.

\clearpage
\section{Orthogonal Security, Privacy, Management and Governance Requirements}
Explainability cannot be isolated from the other cross-cutting planes in Fig.~\ref{fig:ioxt}. An unprotected trace can be manipulated; an unrestricted trace can expose sensitive data; an unmanaged trace can become stale; and a trace without governance may be unusable for accountability.

\paragraph{Security and privacy.} IoT security and privacy challenges include identity, authentication, confidentiality, data protection, safety and misuse of pervasive sensing \citep{Atlam2020,karale2021challenges,abobakr2017iot}. IoXT therefore requires integrity-protected provenance, authenticated identities, selective disclosure, retention controls and stakeholder-specific views.

\paragraph{Management and orchestration.} IoT management addresses monitoring, configuration, fault handling, resource constraints, heterogeneous devices and lifecycle interoperability \citep{aboubakar2022review,sinche2019survey}. Orchestration adds dynamic placement, edge/cloud coordination, QoS, failover and service movement \citep{wen2017fog,shahid2026iot}. IoXT requires explanation continuity to survive these changes.

\paragraph{Governance and compliance.} IoT governance literature identifies legitimacy, transparency, accountability, privacy, interoperability and oversight as system-wide concerns \citep{weber2013internet}. Ethical and transparency work similarly emphasizes consent, information security, physical safety, user awareness and contingency response \citep{allhoff2018internet,baldini2018ethical,wachter2018gdpr}. IoXT adds the claim that the evidence needed for these governance functions should be generated by the architecture itself. Regulatory and risk-management frameworks can provide external assurance requirements, but they should not define the research paradigm \citep{eu2024aiact,nist2023airmf}.

\clearpage
\section{Research Agenda}
\paragraph{Streaming explainability and overhead.} Identity, timestamps, lineage, integrity fields and incident retention consume bandwidth, memory, storage and energy. Research is needed on compact encodings, references, selective logging, aggregation, sampling, edge summarization and adaptive explanation depth, particularly for resource-constrained XAI settings \citep{jagatheesaperumal2022explainable,moss2025explainable,mohammad2025strategies}.

\paragraph{Temporal correctness.} Distributed clocks are imperfect. Clock-quality metadata, uncertainty bounds, sequence numbers and causal ordering should be investigated so that a trace communicates uncertainty rather than implying false temporal precision.

\paragraph{Cross-layer semantic preservation.} Different layers expose different abstractions and lifetimes. Research is needed on how sensor context, application semantics, network events, preprocessing and model decisions should be correlated without creating unmanageable coupling. The cross-layer IoMT results of \citet{georgiades2025explainable} provide one concrete demonstration of the value of retaining such semantics.

\paragraph{Protocol interoperability.} Future work should compare protocol extensions, metadata profiles, gateway rules and new lightweight mechanisms against common XTP requirements. The goal is semantic continuity across heterogeneous transports, not a universal replacement protocol.

\paragraph{Dynamic intelligence and drift.} The trace must survive model updates, federated/distributed learning, mobility, failover and relocation across edge/fog/cloud tiers. Explanations should be anchored to the version and data distribution active at event time \citep{lu2018learning,hu2024explainable}.

\paragraph{Secure provenance.} An adversary can manipulate not only data and predictions but also provenance and explanations. Tamper-evident records, attestation, access control, privacy-preserving provenance and selective evidence export are therefore central research problems \citep{hu2020survey,w3c2013provdm}.

\paragraph{Human-facing explanation.} Engineers, operators, end users, safety officers and auditors require different views. Human-centric XAI provides an important basis for evaluating whether explanations are understandable and actionable \citep{ambritta2024explainable}. IoXT must derive these views from one faithful underlying trace.

\paragraph{System-level evaluation.} IoXT should be evaluated beyond offline XAI benchmarks. Candidate metrics include trace completeness, orphan-actuation rate, cross-layer coverage, temporal consistency, evidence age, semantic preservation, provenance integrity, reconstruction latency, resource overhead and correctness of trace-back/trace-forward. Offline XAI metrics remain useful for the model-level component, but they do not test whether an explanation survives a changing distributed system \citep{kok2023explainable}.

\clearpage
\section{Why the Term IoXT Is Useful}
A new term is justified only if it creates a useful research boundary. IoXT changes the \emph{unit of explanation} from an isolated model to a cyber-physical trajectory; the \emph{evidence} from only model inputs to both data and their communication/transformation history; the \emph{scope} from one analytical layer to cross-layer relationships; the \emph{time of explanation} from predominantly retrospective analysis to streaming and incident-time evidence; the \emph{network role} from transporting model inputs to preserving explanation semantics; and the \emph{engineering status} of explainability from an optional analytics component to a cross-cutting design property.

This distinction matters because IoT links informational behaviour to physical consequences. Existing XAI-for-IoT research remains foundational \citep{kok2023explainable,jagatheesaperumal2022explainable,moss2025explainable,mohammad2025strategies}; IoXT specifies the architectural conditions under which any model-level explanation can remain temporally correct, communication-aware, cross-layer, trackable, traceable and operationally useful across networked Things.

\clearpage
\section{Conclusion}
This paper introduced the \textbf{Internet of Explainable Things (IoXT)} as a system-level paradigm for explainability in intelligent IoT. The central observation is that IoT differs from many conventional XAI settings because evidence is often born in the physical world, communicated through heterogeneous layered networks, transformed by distributed computing, interpreted by AI and policy logic, and, in many high-consequence deployments, converted into actuator actions. The explanation problem therefore concerns both the data and the communication of the data, as well as the intelligence, decision and action. When an action changes the environment, subsequent readings can themselves be consequences of the earlier decision, making event-time traceability even more important.

IoXT consequently argues for \textbf{explainability-by-design and cross-layer explainability}. Relevant Things and events must be identifiable and addressable; observations, transformations, decisions, actions and outcomes must be timestamped and linked; compact evidence must be available during live operation; the system must support trace-back and trace-forward; and missing evidence must be explicit rather than silently ignored. XTP is proposed as a category of mechanisms that can carry these semantics rather than as a fixed protocol.

The defining IoXT requirement can be stated succinctly:
\begin{quote}
\textbf{Every consequential intelligent IoT action should be explainable and traceable, at the time it occurs, back to the timestamped sensed evidence, communication/transformation history and contemporaneous system state that produced it; questionable source evidence should likewise be traceable forward to the decisions and physical actions it may have influenced.}
\end{quote}
This requirement is why IoXT is proposed not merely as XAI applied to IoT, but as a system-level paradigm, design discipline and protocol-design problem for trustworthy networked intelligence and cyber-physical action.

\appendix
\clearpage
\section{Illustrative Design Examples (Non-Normative)}
The main paper defines requirements and design principles. This appendix deliberately separates possible realizations from the IoXT definition.

\subsection{Addressable Explanation Entities}
One possible implementation is to assign stable, namespaced identifiers to explanation-relevant entities and events. These identifiers may be URIs, UUIDs, object identifiers, cryptographic identifiers or domain-specific names. A machine-readable Thing description can advertise which sensor channels, actions, versions and trace capabilities are available \citep{kaebisch2023web}. The important requirement is not the identifier syntax; it is the ability to resolve the entity that participated in a historical trace.

\subsection{Distributed Explanation State}
Explanation state need not reside in one database. A constrained sensor may retain recent sequence numbers, hashes and calibration references; a gateway may retain short-term lineage; an edge node may retain transformation state; a cloud service may retain model and decision explanations; and an incident service may retain longer-lived bundles. This distributed design is conceptually similar to established management practice in which structured state is exposed remotely, but IoXT does not inherit SNMP naming or information-base structures \citep{rfc3411}.

\subsection{Example Minimum Record}
Equation~\eqref{eq:envelope} can be instantiated as the following non-normative logical record:
\begin{verbatim}
entity_id       = sensor/site7/temp/03
observed_at     = 2026-08-08T17:42:15.238+03:00
trace_id        = 7fa3...e91
parents         = []
version         = calibration:v12
context         = unit:C, location:zone4
quality         = confidence:0.97, clock_uncertainty_ms:2
integrity       = hash/signature/reference
\end{verbatim}
A downstream transformation, model inference, decision and actuator event would create corresponding records linked through the same trace identifier and parent relationships. Rich LIME, SHAP, SENN, C-SENN or other model explanations can be stored separately and referenced by the model event.

\subsection{Example XTP Operation Set}
An XTP realization could expose operations equivalent to:
\begin{itemize}
\item \textbf{PUBLISH(record)} -- emit a timestamped trace/explanation record;
\item \textbf{LINK(child,parent)} -- bind an output to source evidence;
\item \textbf{QUERY(id,trace,time)} -- reconstruct state by entity, trace or time;
\item \textbf{SUBSCRIBE(scope)} -- receive live explanation/trace events; and
\item \textbf{FREEZE/ATTEST(window)} -- preserve and integrity-protect an incident interval.
\end{itemize}
These operations could be mapped to current protocols or implemented separately. For example, MQTT properties/topics can carry correlation and metadata in a publish--subscribe deployment \citep{mqtt5}; CoAP resources and options can expose constrained trace resources \citep{rfc7252}; and management-style queries can inspire remote access to explanation state \citep{rfc3411}. These are examples, not prescribed bindings.

\subsection{Example Incident Reconstruction}
Assume actuator action $a_t$ closes a valve and an incident occurs. A reverse query resolves the authorizing decision $d_t$, the model inference $m_t$, the transformed input $z_t$, and sensor observations $s_1,s_2,s_3$. The system returns $\mathcal{S}(a_t)=\{s_1,s_2,s_3\}$, their timestamps/calibration states, the communication transformations, the model version and model-level explanation, and the execution result. If $s_2$ is later found faulty, Eq.~\eqref{eq:impactset} is used to identify other actions reachable from $s_2$ during the affected interval. This does not prove that $s_2$ caused the incident; it identifies the evidence path that subsequent causal or forensic analysis must inspect.

\clearpage
\section{Illustrative Conformance Profile (Non-Normative)}
\begin{longtable}{p{0.25\linewidth} p{0.68\linewidth}}
\caption{Illustrative minimum checks for a future IoXT profile.}\label{tab:appendixconformance}\\
\toprule
\textbf{Capability} & \textbf{Illustrative check} \\
\midrule
\endfirsthead
\toprule
\textbf{Capability} & \textbf{Illustrative check} \\
\midrule
\endhead
Identity & Each declared sensor channel, model/policy instance, decision and actuator event is resolvable by a stable logical identity. \\
Time & Explanation-relevant events expose event time and clock uncertainty/quality where required. \\
Version state & Calibration, configuration, firmware, model and policy versions are recoverable for the incident interval. \\
Lineage & Parent/input references permit traversal from sensor evidence through transformations and intelligence to actuation. \\
Streaming & Compact provenance or references are produced while the system is operating. \\
Cross-layer evidence & Required sensing, communication, model, decision and actuation layers are represented, yielding $C_{\mathrm{layer}}=1$ for the declared profile. \\
No orphan actuation & Critical actions satisfy $C_{\mathrm{trace}}=1$ or carry an explicit degraded/non-conformant state. \\
Temporal consistency & The action trace satisfies the declared temporal tolerance and reports any violations used in Eq.~\eqref{eq:ctime}. \\
Protocol continuity & Required semantics survive each gateway/protocol mapping, or missing fields are explicitly reported through Eq.~\eqref{eq:csem}. \\
Incident capture & The relevant time window can be frozen/attested without rewriting the historical record. \\
Security/privacy & Trace integrity, authorization, confidentiality/selective disclosure and retention are enforced according to the domain profile. \\
Traversal & Both reverse action-to-evidence and forward source-to-impact queries are demonstrated in conformance testing. \\
\bottomrule
\end{longtable}

\clearpage
\bibliographystyle{unsrtnat}
\bibliography{references}

\end{document}